\PassOptionsToPackage{table}{xcolor}
\documentclass[sigconf]{acmart}

\setcitestyle{square,sort,comma,numbers}

\usepackage{CJKutf8}

\usepackage[utf8]{inputenc} 
\usepackage{hyperref}       
\usepackage{url}            
\usepackage{booktabs}       
\usepackage{amsfonts}       
\usepackage[hide]{notes-alt}
\usepackage{nicefrac}       
\usepackage{microtype}      
\usepackage{xcolor} 
\usepackage[table]{xcolor}
\usepackage{fontawesome5}
\usepackage{colortbl}        

\usepackage{amstext}
\usepackage{amsmath}
\usepackage{enumitem}
\usepackage{algorithm}
\usepackage{algpseudocode}
\usepackage{dblfloatfix}
\usepackage{multirow}
\usepackage{wrapfig}
\usepackage{tikz}
\usetikzlibrary{shapes.geometric, arrows, positioning, shadows}
\usepackage{pgfplots}
\usepackage{subcaption}
\usepackage{pgfplotstable}
\usetikzlibrary{patterns}

\usepackage{xcolor}

\usepackage{subcaption}

\usetikzlibrary{
    positioning,
    arrows,
    calc,
    shapes.callouts,
    decorations.pathmorphing,
    decorations.text,
    fit,
    backgrounds,
}
\usepgfplotslibrary{statistics}
\pgfplotsset{compat=1.15}
\definecolor{c1}{HTML}{e41a1c}
\definecolor{c2}{HTML}{377eb8}
\definecolor{c3}{HTML}{4daf4a}
\definecolor{c4}{HTML}{984ea3}
\definecolor{reformoptgray}{gray}{0.93}

\newcommand{\idnt}{\phantom{ - }}
\newcommand{\sys}[1]{\textsc{Gar}\def\temp{#1}\ifx\temp\empty{}\else\raisebox{-.4ex}{\scriptsize #1}\fi}

\newcommand{\sgar}[1]{\textsc{Sgar}\def\temp{#1}\ifx\temp\empty{}\else\raisebox{-.4ex}{\scriptsize#1}\fi}

\newcommand{\sysbm}[1]{\sys{}${}_{BM25}$}
\newcommand{\quamsys}[1]{\textsc{Quam}\def\temp{#1}\ifx\temp\empty{}\else\raisebox{-.4ex}{\scriptsize #1}\fi}

\newcommand{\cerberussys}[1]{\textsc{ORE}\def\temp{#1}\ifx\temp\empty{}\else\raisebox{-.4ex}{\scriptsize #1}\fi}
\newcommand{\cerberuhalfssys}[1]{\textsc{CerberusHalf}\def\temp{#1}\ifx\temp\empty{}\else\raisebox{-.4ex}{\scriptsize #1}\fi}

\newcommand{\approach}{\textsc{ReformIR}}

\newcommand{\argmaxm}[1]{%
  \ifthenelse{\isempty{#1}}%
    {\overset{m}{\argmax}}
    {\underset{#1}{\overset{m}{\argmax}}\, }
}
\newcommand{\argminm}[1]{%
  \ifthenelse{\isempty{#1}}%
    {\overset{m}{\argmin}}
    {\underset{#1}{\overset{m}{\argmin}}\, }
}
\usepackage{pifont}

\newcommand{\sgarsys}[1]{\textsc{SlideGar}\def\temp{#1}\ifx\temp\empty{}\else\raisebox{-.4ex}{\scriptsize#1}\fi}

\author{Venktesh V}
\orcid{0000-0001-5885-2175}
\affiliation{%
    \institution{DSV, Stockholm University}
    \city{Stockholm}
    \country{Sweden}
}
\email{venktesh.viswanathan@dsv.su.se}

\author{Mandeep Rathee}
\orcid{0000-0002-7339-8457}
\affiliation{%
  \institution{L3S Research Center}
  \city{Hannover}
  \country{Germany}  
}
\email{rathee@l3s.de}

\author{Avishek Anand}
\orcid{0000-0002-0163-0739}
\affiliation{%
    \institution{Delft University of Technology (EEMCS/EWI, TU~Delft)}
    \city{Delft}
    \country{The Netherlands}
}
\email{avishek.anand@tudelft.nl}

\makeatother

\begin{document}

\title{Adaptive Query Optimization and Document Relevance Estimation through Ranker Feedback}

\title{When More Reformulations Hurt: Avoiding Drift using Ranker Feedback}


\begin{abstract}

Modern retrieval pipelines increasingly rely on query reformulation and neural reranking to improve effectiveness, but this comes at a significant computational cost and introduces a fundamental tradeoff between \textit{recall} and \textit{query drift}. 
Generating many reformulated queries can substantially increase recall, yet na\"ively merging or exhaustively reranking their results is prohibitively expensive. 
In this work, we argue that the core challenge is not reformulation generation itself, but the adaptive selection of reformulations and their retrieved documents under a strict inference budget.

We propose \approach{}, a budget-aware retrieval framework that treats query reformulations as first-class features and performs online relevance estimation using a strong reranker as a teacher. Given multiple reformulated queries, \approach{} constructs a large candidate pool and learns a lightweight surrogate model that estimates document utility from reformulation-specific retrieval signals. 
Under a fixed reranking budget, the surrogate adaptively prioritizes both reformulations and documents, selectively querying a teacher reranker anchored to the original query. This process increases recall while actively suppressing drift through online feature selection over reformulations.
We conduct extensive experiments on the MSMARCO passage corpora and TREC Deep Learning benchmarks (DL19–DL22), evaluating effectiveness under realistic reranking budgets. 
Our results show that \approach{} consistently outperforms existing reformulation strategies, particularly as the number of reformulations increases, where prior methods suffer from severe quality degradation due to drift. 
Our findings also suggest a shift in retrieval system design: rather than using large language models as rerankers, their capacity is more effectively leveraged in the reformulation stage with feedback-driven optimization. Our code: 

\vspace{0.5em}
\hspace{2.0em}\includegraphics[width=1.25em,height=1.25em]{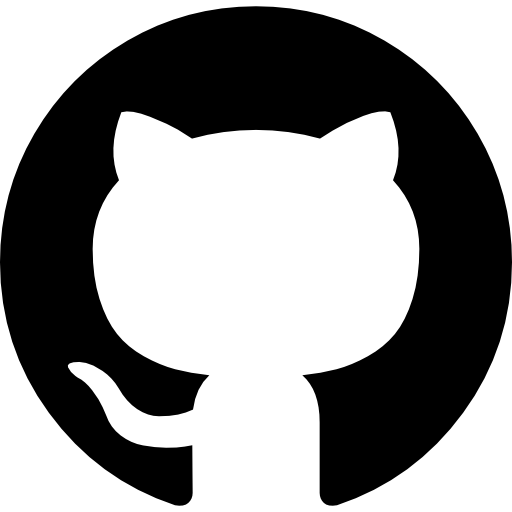}\hspace{.3em}
\parbox[c]{\columnwidth}
{
    \vspace{-.55em}
    \href{https://github.com/VenkteshV/ReformIR}{\nolinkurl{https://github.com/VenkteshV/ReformIR}}
}
\vspace{-1.5em}
\end{abstract}


\maketitle

\section{Introduction}

Query expansion is one of the fundamental tasks in document ranking and has been studied for a long time in the field~\cite{manning,carpineto2012survey}.
Long-standing work has shown that reformulating a user query, by adding,
removing, or reweighting terms, is often essential to overcome vocabulary mismatch between queries and documents using user or pseudo-relevance feedback~\cite{rocchio1971relevance,rm3,carpineto2012survey}.
At the same time, this line of work also identified a fundamental risk: indiscriminate expansion can introduce query drift, where added terms shift the query away from the user’s original intent and degrade performance at lower depths \cite{manning}.

Instead of relying solely on corpus statistics, modern approaches use LLMs to
generate reformulated queries, expansion terms, or even pseudo-documents based on their parametric memory and their improved language understanding.
Methods such as Query2Doc \cite{wang2023query2doc} generate synthetic documents from the query, while GenQR and its variants \cite{genqr,genqr_ensemble} directly prompt LLMs to rewrite or expand queries.
More recent work, such as QA-Expand \cite{seo2025qa}, introduces diversity through question--answer generation, using LLMs to synthesize multiple sub-questions and answers as expansion signals.

\begin{figure}[]
    \centering
    \begin{subfigure}{0.48\columnwidth}
        \centering
 \begin{tikzpicture}
		\begin{axis}[
			width = 1.25\linewidth,
			height =1.25\linewidth,
			major x tick style = transparent,
			grid = major,
		    grid style = {dashed, gray!20},
			xlabel = {\# of reformulations},
			ylabel = {},
			title={Recall@100},
            title style={yshift=-1.5ex}, 
            symbolic x coords={0,3,5,10,15,25,50},
            xtick={0,3,5,10,15,25,50},
            xtick distance=20,
            enlarge x limits=0.05,
            xlabel near ticks,
            ylabel near ticks,
            ymin=0.42,
            ymax=0.6,
            every axis y label/.style={at={(-0.18, 0.5)},rotate=90,anchor=near ticklabel},
			]

			

   	\addplot [color=black, line width = 0.8pt] table [x index=0, y index=3, col sep = comma] {fig/varying_n_vs_performance.txt}
    node[pos=0.5, sloped, above] {\small{GenQR}};

	\addplot [color=black, style= dashed, line width = 1.0pt] table [x index=0, y index=1, col sep = comma] {fig/genqr_ensemble_varying_n.txt}
         node[pos=0.5, below, yshift=1pt] {\small{BM25}};
   	\addplot [color=blue, line width = 0.8pt] table [x index=0, y index=4, col sep = comma] {fig/varying_n_vs_performance.txt}
    node[pos=0.4, yshift=5pt, above] { \small{\approach{}}};
   
    \end{axis}
    \end{tikzpicture}    
    \caption{} 
    \label{fig:genqr}
    \end{subfigure} 
    \hfill
    \begin{subfigure}{0.48\columnwidth}
    \centering
    \includegraphics[width=1.1\columnwidth]{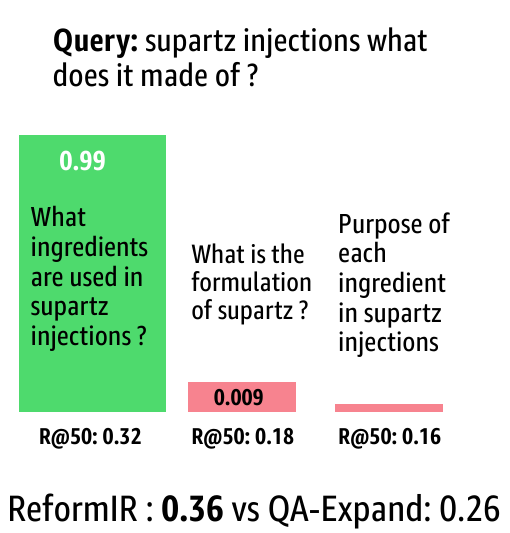}
     \caption{} 
    \label{fig:qual_anaysis}
    \end{subfigure}


      \caption{\textbf{Drift control and learned reformulation weights.}
\textbf{(a)} Recall@100 on TREC DL19 as the number of reformulations increases.
While GenQR suffers from performance degradation as additional reformulations
introduce drift, \approach{} remains stable and consistently improves recall by
selectively prioritizing effective reformulations using ranker feedback.
\textbf{(b)} Reformulation weights learned by \approach{} through online
optimization for a representative query (We also show Recall@50 for individual queries, QA-Expand baseline and \approach{}).}\label{fig:scaling_exp}  
    \Description{}
\end{figure}
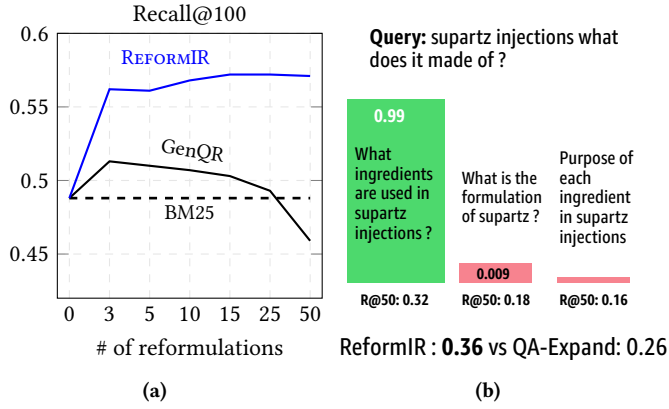

Despite their empirical success, existing LLM-based reformulation methods largely share a common limitation.
They are potentially effective at the \emph{recall-increasing step} of retrieval, retrieving documents that would be missed by a single query, but remain weak at the \emph{drift-reduction step}.
Generated reformulations are typically combined using heuristic rules, fixed
interpolation weights, or post-hoc filtering, without being explicitly guided by retrieval feedback.
As a result, irrelevant or weak reformulations often contribute noise, forcing systems to conservatively down-weight expansion or accept increased drift.
We argue that in this sense, many modern approaches \textit{inherit} the classical query drift problem, but without the explicit feedback mechanisms that earlier relevance models used to control it.

In this work, we introduce \approach{}, a retrieval framework that uses LLMs
exclusively as \emph{low-cost reformulation generators} and employs a budgeted reranker as a feedback mechanism to re-weight reformulations.
\approach{} generates multiple reformulations, retrieves documents independently for each reformulation to maximize recall, and then learns, under a strict budget, which reformulations are useful.
Inspired from existing works on online feature attributions in explainable IR~\cite{singh2020model,lyu2023listwise}, reformulations are treated as features in a surrogate relevance model.
We then employ a bandit-style optimization loop that allocates a limited number of reranker calls to progressively focus on documents retrieved by promising reformulations.
Importantly, the reranker always evaluates relevance with respect to the original query, anchoring optimization, and actively suppressing query drift.

This design yields several advantages. First, it decouples query understanding from expensive document-level reasoning. 
LLMs are used where they are strongest and cheapest: for understanding and rephrasing queries, rather than where they are most costly.
Second, because reformulation operates in a short-context regime, \approach{} naturally supports the use of smaller language models, enabling further reductions in inference cost.
Third, the framework is generator-agnostic and can be layered like an adapter on top of any reformulation method; our experiments show consistent improvements across a wide range of generators.
Finally, by learning explicit weights over reformulations, \approach{} provides interpretability, revealing the aspects of a query driving relevance.

Figure~\ref{fig:genqr} provides a concrete illustration of the query drift problem and how \approach{} addresses it.
The figure plots Recall@100 as a function of the number of reformulations for a classical LLM-based reformulation approach (GenQR) and its counterpart augmented with \approach{}.
As the number of reformulations increases, the classical approach exhibits a clear degradation in performance: while additional reformulations initially improve recall, further expansions introduce irrelevant reformulations that dominate the retrieval pool, leading to drift and a steady decline in effectiveness.
In contrast, \approach{} remains robust as the number of reformulations grows.
By using ranker feedback to learn which reformulations are actually useful and suppressing those that induce drift, \approach{} maintains stable and consistently higher recall even under aggressive reformulation scaling.
This highlights a key distinction between naive expansion and adaptive optimization: simply adding more reformulations amplifies noise, whereas \approach{} actively controls drift by anchoring relevance feedback to the original query.

Our extensive experiments on TREC Deep Learning benchmarks demonstrate that \approach{} functions as a training-free adapter that consistently reduces query drift across reformulation methods, yielding gains of up to \textbf{33.48\%}.
While classical reformulation approaches degrade as the number of reformulations increases, \approach{} remains stable and continues to improve performance even.
These benefits hold across a diverse set of reformulation generators and model scales, while incurring only marginal additional latency compared to classical approaches and remaining \textbf{3.3$\times$--4.5$\times$} more efficient than LLM-based reranking.

This paper makes the following key contributions:

\begin{itemize}
    \item We introduce \approach{}, a novel retrieval framework based on
    bandit-style optimization that actively helps suppress query
    drift using reranker feedback.

    \item We provide extensive empirical evidence showing that \approach{} consistently improves retrieval effectiveness across different reformulation baselines serving as a training-free adapter.

    \item We demonstrate how LLMs should be pragmatically
    integrated into retrieval systems, combined with lightweight, feedback-driven optimization, resulting in superior
    effectiveness, lower latency, and interpretability compared to
    downstream LLM-based reranking.

\end{itemize}

\section{Related Work}
\label{sec:rel_work}

\todo{add some more papers structure it - Venky}
 The efficient first stage retrievers, like BM25, only focus on lexical matching and perform poor when there is a vocabulary mismatch, especially for web queries which are usually short and ambiguous. Towards this end, several query expansion and reformulation techniques have been proposed to mitigate this issue.


\subsection{Traditional Query Reformulation}

Traditional approaches to query reformulation primarily rely on test-time expansion strategies to enhance lexical overlap. Pseudo-Relevance Feedback (PRF) methods, such as RM3~\cite{rm3}, assume that top-ranked documents in an initial retrieval pass contain relevant terms; these terms are then extracted to augment the original query. Similarly, the Rocchio algorithm~\cite{rocchio1971relevance} provides a formal framework for query vector modification by incorporating weighted feedback from both relevant and non-relevant documents.

Further refinements introduced probabilistic and information-theoretic perspectives. KL Expansion~\cite{zhai2001model} utilizes language modeling to optimize interpolation coefficients between the query and expansion terms, while Relevance Modeling (RM)~\cite{lavrenko2001rm} estimates a query-specific language model from top-retrieved passages. Beyond simple term frequencies, Latent Concept Expansion (LCE)~\cite{metzler2007latent} uses Markov Random Fields to identify higher-level concepts, and Local Context Analysis (LCA)~\cite{xu1996query} examines co-occurrence patterns within the local retrieval context. These methods collectively establish the foundation for modern test-time query reformulation. Though these methods use initially retrieved documents based on lexical overlap, this overlap is still bounded by the original query. Moreover, these methods employ heuristic strategies for combining reformulations and fail to identify and mitigate semantic drift. 



\subsection{Generative Query Reformulation} 
More recently, powerful generative language models have been found to generate well-formed variants of the original query \cite{genqr,jagerman2023query,song2024survey,anand2023query:age:vision}. 
Several methods have been proposed to employ off-the-shelf LLMs, which can be broadly divided into the following classes: rephrasing-based \cite{genqr,genqr_ensemble}, pseudo-relevance feedback based \cite{genqr_ensemble}, pseudo-document generation \cite{wang2023query2doc}, and question answer generation \cite{seo2025qa} based approaches. Since the web style queries are usually shorter and face the vocabulary mismatch problem, LLM reformulated queries can help find further relevant documents.     

GenQR \cite{genqr} employs generative LLMs to rephrase the original query to generate reformulations, but employs heuristic filtering mechanisms to prevent topic drift with respect to the original query. GenQREnsemble \cite{genqr_ensemble} argues that reformulations based on a single instruction may lead to worse performance and proposes ensembling of reformulations from multiple instructions to the LLM in a zero-shot setting. Generative Relevance Feedback (GRF)~\cite{mackie2023generative,mackie2023grm,mackie2023generativeall} 
generate several reformulations, for example, by generating keywords, entities, documents, an essay, or a news article. Query2Doc~\cite{wang2023query2doc} employs zero-shot / few-shot by providing query-document pairs and generates a document for a given query.  Finally, the generated document is concatenated with the original query. Methods like QueryExpansion~\cite{jagerman2023query} expand the query in a few-shot or CoT (Chain-of-Thoughts) setup ~\cite{wei2022chain}. HyDE~\cite{gao-etal-2023-precise}, which generates hypothetical relevant documents (can be factually incorrect), using parametric knowledge of LLMs, for a given query and uses their embeddings as enhanced query representations for dense retrieval. However, these hypothetical documents might lead to concept drift in the query, and the retrieval system ends up with noisy documents. 


While effective, these "zero-shot" generative methods can introduce concept drift if the LLM generates factually incorrect text. Recent hybrid approaches like Mill~\cite{jia-etal-2024-mill}, BlendFilter~\cite{wang-etal-2024-blendfilter}, MuGI~\cite{zhang-etal-2024-exploring-best}, and ReAL~\cite{chen-etal-2025-terms}, and pseudo-relevance feedback-based methods like GenQREnsemblePRF \cite{genqr_ensemble} integrate initial retrieved documents into the generation process to try and mitigate this issue. However, \cite{wang2023query2doc} observes that when the quality of the original query and its retrieved results are poor, it causes further drift compared to zero-shot approaches. Furthermore, models like GAR~\cite{mao-etal-2021-generation} and LameR~\cite{shen-etal-2024-retrieval} iteratively refine the query based on the LLM's own generated answers, ensuring better alignment with the final task.

For purposes of retrieval, the existing works combine the initial query with reformulated queries in several ways. Primarily, most works adapt a heuristic approach by concatenating the initial query or repetitions of the initial query with the reformulated queries to perform retrieval \cite{seo2025qa,genqr,wang2023query2doc,genqr_ensemble}
. However, this approach also incorporates reformulations with drift, which causes the retrieval of irrelevant documents, impacting performance. Alternatively, individual reformulated queries are used to retrieve documents, which are then combined using approaches like RRF \cite{rrf} in approaches like \cite{seo2025qa,query_expand_cross_enc,dhole2024generativequeryreformulationusing}. However, RRF considers all queries to be equal. Although it is intended to capture multiple facets of queries, it often combines redundant or low-quality candidates as observed in \cite{seo2025qa}, causing deterioration in performance. Our work focuses on a principled approach to adaptively prioritize high-quality queries and documents using a feedback mechanism from the reranker. 

While prior works like ODIS \cite{odis}, explore distilling re-ranker to a smaller lexical scoring function, they require scoring of all documents from first-stage retrieval for a query to enable the distillation. This is followed by the selection of documents from the whole corpus using the distilled model with another round of re-ranking by the cross-encoder model.  However, our approach focuses on adaptively prioritizing promising reformulations and also prioritizes corresponding documents. Our approach also relies only on linear combination of simple features for the linear surrogate and obviates need for re-ranking all retrieved documents corresponding to reformulations.

While other line of works, such as DeepRetrieval~\cite{jiang2025deepretrieval}, where LLM is fine-tuned using RL for the query reformulation task exists, we focus only on \textbf{training-free} query reformulation methods. 
\begin{figure*}
\centering

\tikzset{
    intent/.style={
        align=left,
        text width=10em,
        inner sep=0.8em,
        fill=gray!15,
        rounded corners,
        draw=black,
        label={
            [fill=black, rounded corners, font=\sffamily\scriptsize, text=white,
             anchor=west, xshift=-5em, inner sep=0.25em]above:#1
        },
    },
    annotation/.style={
        draw=black,
        fill=white,
        rounded corners=2pt,
        inner sep=0.25em,
        font=\scriptsize\itshape,
        execute at begin node={$\,\psi(Q,d)\,$},
    },
    desc/.style={
        align=right,
        text width=11em,
    },
    identifier/.style={
        font=\small,
    },
    stepnum/.style={
        circle,
        fill=black,
        text=white,
        font=\small\bfseries,
        inner sep=0.2em,
        outer sep=0.2em,
    }
}

\begin{tikzpicture}

    \node[
        draw=gray,
        rounded corners,
        inner sep=0.75em,
    ] (Q) {what is the sacraments of service in the catholic church? \qquad\qquad \faIcon{search}};

    \node[
        desc,
        left=1.25 of Q,
    ] (Q-d) {\textbf{Query}};
    \node[
        stepnum,
        right=0.25 of Q-d,
    ] (Q-d-s) {};

    \node[
        identifier,
        above=0.25 of Q.north east,
        anchor=east,
    ] {$$Q$$};

    \node[
        intent={Reformulation},
        below=1.75 of Q.west,
        anchor=west,
    ] (I-1) {What are the two sacraments of service in the Catholic Church?};
    \node[
        identifier,
        above=0.25 of I-1.north east,
        anchor=east,
    ] {$Q_{1}$};
    \draw[-latex] (Q) to[
        out=275,
        in=85,
    ] (I-1);

    \node[
        intent={Reformulation},
        right=0.5 of I-1,
    ] (I-2) {Purpose of the sacraments of service in the Catholic faith?};
    \node[
        identifier,
        above=0.25 of I-2.north east,
        anchor=east,
    ] {$Q_{2}$};
    \draw[-latex] (Q) to[
        out=340,
        in=100,
    ] (I-2);

    \node[
        intent={Reformulation},
        right=0.5 of I-2,
    ] (I-3) {Difference of sacraments of service from other catholic sacraments? };   
    \node[
        identifier,
        above=0.25 of I-3.north east,
        anchor=east,
    ] {$Q_{3}$};

    \coordinate (qCtrl)  at ($(Q.east)  + (.4, .4)$);
    \coordinate (i3Ctrl) at ($(I-3.north)+ (0.0, .2)$);

    \draw[-latex]
        (Q.east)
        .. controls (qCtrl) and (i3Ctrl)
        .. (I-3.north);

    \node[
        desc,
    ] (I-d) at (Q-d |- I-1) {\textbf{Reformulator $\mathcal{R}$}\par
\footnotesize{Generate $m$ reformulations:
$\mathcal{Q}=\{Q_1,\ldots,Q_m\}$.}};
    \node[
        stepnum,
    ] (I-d-s) at (Q-d-s |- I-d) {1};

    \foreach \p [count=\idx] in {43, 12, 3, 67, 90, 56}
    {
        \coordinate (L) at ($(I-1.west) + (0.75, 0)$);
        \coordinate (R) at ($(I-3.east) + (-0.75, 0)$);
        \coordinate (C) at ($(L)!1/5*(\idx-1)!(R) + (0, -2.5)$);
        \node[
            outer sep=0.25em,
        ] (P-\idx) at (C) {\Huge \faIcon{file-alt}};
        \node[
            identifier,
            right=-0.15 of P-\idx,
        ] (P-i-\idx) {$d_{\p}$};
    }

    \coordinate (Temp) at ($(P-1)!0.5!(P-i-1)$);
    \node[
        desc,
    ] (P-d) at (Temp -| I-d) {\textbf{Online feat. selection}\\ \footnotesize{updating surrogate model from teacher ranking } $s(d; \mathbf{w}) = \mathbf{w}^\top \mathbf{x}_d$};
    \node[
        stepnum,
    ] (P-d-s) at (Q-d-s |- P-d) {3};

    \node[
        desc,
    ] (C-d) at ($(I-d)!0.5!(P-d)$) {\textbf{Reranking}\\ \footnotesize{using Teacher with budget $\tau$}};
    \node[
        stepnum,
    ] (C-d-s) at (Q-d-s |- C-d) {2};

    \draw[-latex] (I-1) to[
        out=250,
        in=85,
    ] node[annotation] {} (P-1);

    \draw[-latex] (I-1) to[
        out=300,
        in=140,
    ] node[annotation, pos=0.2] {} (P-3);

    \draw[-latex] (I-1) to[
        out=320,
        in=150,
    ] (P-5); 

    \draw[-latex] (I-2) to[
        out=230,
        in=70,
    ] node[annotation, pos=0.4] {} (P-2);

    \draw[-latex] (I-2) to[
        out=270,
        in=80,
    ] node[annotation, pos=0.4] {} (P-3);

    \draw[-latex] (I-3) to[
        out=240,
        in=70,
    ] node[annotation] {} (P-4);

    \draw[-latex] (I-3) to[
        out=260,
        in=80,
    ] (P-5); 

    \draw[-latex] (I-3) to[
        out=290,
        in=90,
    ] (P-6); 

    \foreach \I/\p [count=\idx] in {{20}/43, {21}/12, {20}/3, {22}/67, {20}/90, {22}/56}
    {
        \foreach \i [count=\n] in \I
        {
            \coordinate (T) at ($(P-\idx) + (0, -0.5*\n-0.6)$);
            \node (T-\idx-\n) at (T) {$($Q$, d_{\p})$};
        }
        \draw[-latex] (P-\idx) -- (T-\idx-1);
    }

    \coordinate (Temp) at ($(T-3-1)!0.5!(T-3-1)$);
    \node[
        desc,
    ] (T-d) at (P-d |- Temp) {\textbf{Final ranking} };

    \node[
        stepnum
    ] (END) at (Q-d-s |- T-d) {}; 

    \draw[
        -latex,
        dashed,
    ] (Q-d-s) -- (I-d-s) -- (C-d-s) -- (END);

    \draw[
        -latex,
        dashed,
    ] (P-d-s) -- (C-d-s);

    \draw[
        -latex,
        dashed,
    ] (C-d-s) -- (P-d-s);

\end{tikzpicture}
\caption{A broad overview of proposed approach \approach{}. The original query is reformulated to several queries as shown. This is followed by BM25 retrieval to form candidate pool as detailed in Section \ref{sec:method}. Then  Steps 2-3 adaptively prioritize documents using relevance estimated by surrogate for sampling ranker feedback and performs updates to the surrogate are iterative as detailed in Algorithm \ref{algo:cerberus}.   
Final ranking shows that reformulated queries $Q_1$ and $Q_2$ are more closer semantically to the original query and hence they are upweighted and their documents are retrieved and  ranked higher. $Q_3$ focuses on ``difference of sacraments" and drifts from original meaning of the query which results in downweighting it and corresponding documents.}
\label{fig:method}
\end{figure*}
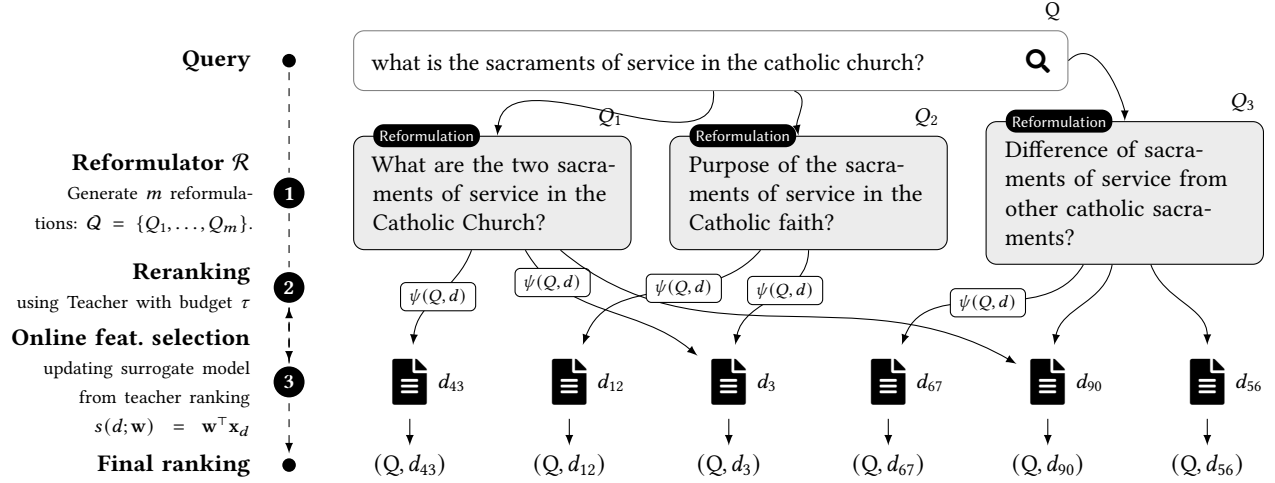



\section{Our Approach \approach{}}
\label{sec:method}
In this section, we present our approach in detail, addressing the challenges of using multiple query reformulations for retrieval. Our aim is to achieve the high recall afforded by multiple reformulations with a low budget on reranker evaluations by avoiding query drift. An overview of the pipeline is shown in Figure~\ref{fig:method}.

\subsection{Preliminaries and Problem Setup}


Let $Q$ be the original user query expressing an information need, and let $\mathcal{D}$ be a large document collection. Modern retrieval systems often generate multiple reformulations of $Q$ (alternative phrasings, expansions, etc.) to improve recall, especially for complex or ambiguous queries. The benefit is that different reformulations can retrieve different relevant documents. The drawback is a well-known issue of \textit{query drift}: some reformulations might stray from the user’s intent, retrieving off-target documents~\cite{turtle1995query}. 


Hence, apart from generating a wide range of reformulated queries; the key challenge is deciding which reformulations (and their results) to prioritize and which to down-weight or ignore. We tackle this by treating reformulations as latent features of the retrieval process and using a strong reranker, anchored to the original query, as a guide (or “teacher”) to dynamically select and weight reformulations under a strict budget of reranker evaluations.

\subsection{Overview of the \approach{} Pipeline}

Our proposed pipeline, called \approach{}, is designed to get the best of both worlds: it leverages multiple query reformulations and employs adaptive selection of reformulation and documents from candidate pool utilizing reranker feedback carefully to minimize drift and achieve higher recall. The pipeline consists of four main stages:

\subsubsection{Reformulations Generation} Produce a set of $m$ reformulated queries from the original query $Q$. These reformulations ${Q_1, Q_2, \ldots, Q_m}$ aim to cover various interpretations or aspects of the query, increasing the chance of retrieving all relevant information.
    
\subsubsection{Candidate Retrieval} Use a first-stage retriever (we use BM25) to execute each reformulated query $Q_i$ and obtain a list of top-$k$ documents $D_i = [{d_{i1}, d_{i2}, \ldots, d_{ik}}]$. The initial results for the original query $Q$ (denoted $R_0$) are also obtained. All retrieved documents are combined into a candidate pool:

    $$
\mathcal{C} \;=\; R_0 \;\cup\; \bigcup_{i=1}^{m} D_i~,$$

where $\mathcal{C}$ may contain duplicate documents if the same document appears in multiple $D_i$ (in practice, such duplicates are de-duplicated in the set union). This union operation ensures maximal recall, any document that is highly ranked for any reformulation or for the original query will be considered. However, $\mathcal{C}$ can be quite large and potentially noisy, since some reformulations might pull in irrelevant content. While we typically cap the depth $k$ of each list (typically 100) to keep $|\mathcal{C}|$ manageable, while still allowing $\mathcal{C}$ to be large enough to include most relevant documents it could still comprise of irrelevant documents from low quality reformulations. Figure~\ref{fig:method} illustrates this multi-reformulation retrieval process.
    
\subsubsection{Surrogate Model (Reformulations as Features)} We embed reformulations into a feature-based surrogate model that can predict document relevance. Instead of deciding upfront which reformulations are useful, we assign each reformulation a weight that will be learned from data. Concretely, for each document $d \in \mathcal{C}$, we construct a feature vector $\vec{x}_d\in\mathcal{R}^{m}$ with one dimension for each reformulated query’s contribution to $d$. 
\vspace{-0.2em}
\begin{equation}
     \vec x_d=\big[
\text{BM25}(Q_1,d),.., \text{BM25}(Q_m,d),
\text{BM25}(Q,d), \text{RM3}(Q',d)
\big]
\label{eq:features}
\end{equation}


    
This way, each reformulation $Q_i$ is represented as a feature indicating how well document $d$ matches that particular query. The original query’s BM25 score is also included as a feature, and we add an RM3-based feature to capture pseudo-relevance feedback signals. In our implementation, $Q’$ for RM3 is an expanded version of the original query $Q$ computed using a small set of documents (\textbf{$S$}) already identified as relevant by the reranker (initially, this could be the top results from $R_0$ \cite{rathee2025quam}, and it can be updated as more feedback is sampled). The RM3 feature for document $d$ is essentially the query-likelihood or relevance model score of $d$ under this expanded query $Q’$, which measures the vocabulary overlap between $d$ and the known relevant set of documents (already ranked). Intuitively, while the BM25 features capture direct term overlap with each query formulation, the RM3 feature captures indirect relevance: it indicates whether $d$ shares terms with other documents that were deemed relevant, thereby serving as a drift-resistant signal, which results in $\vec{x}_d\in\mathcal{R}^{m+2}$.

\subsection{Surrogate Relevance Model}

We denote our surrogate relevance model as a linear function 
$s(d; \vec{w})$ that estimates the relevance of document $d$ given the feature vector:
$s(d; \vec{w}) \;=\; \vec{w}^\top \vec{x}_d~$,
where $\vec{w}$ is a weight vector that the model will learn using reranker feedback. This estimated utility using an inexpensive surrogate is utilized to prioritize documents for sampling reranker feedback as explained in the following section. Note that the re-ranker feedback is sampled with respect to the original query which serves as a reference point.  This also aligns with the notion in query reformulation, to generate query variations that preserve the semantics of the original query \cite{genqr,genqr_ensemble,conquer}. Each component of $\vec{w}$ corresponds to the importance of the associated feature (and thereby the associated reformulation or original query). We first initialize $\vec{w}$ set to some default (e.g., small random) values. Learning $\vec{w}$ is effectively learning which reformulations are useful: if reformulation $Q_i$ tends to retrieve relevant documents, the weight $w_i$ will increase; if $Q_i$ retrieves mostly irrelevant documents, $w_i$ will decrease, potentially becoming very small if that reformulation is consistently misleading.
We frame this \textit{surrogate learning problem} in terms of an \textit{online learning} or bandit problem. 

\vspace{-1em}
\subsection{Modeling documents as multi-arm bandits}

In a multi-armed bandit scenario, each arm is an option that can be sampled for obtaining feedback in the form of rewards from an oracle. In our setting, each document can be cast as an ``arm'' with reformulations implicitly encoded as part of the utility estimation of the arm using a linear surrogate as described above, and the goal is to maximize reward by sequentially selecting arms and learning from feedback. 
In our setting, each candidate document $d \in \mathcal{C}$ (an arm) has an unknown true relevance reward with respect to the original query, which we can only observe by sampling from the expensive reranker (teacher) $\psi$. The feature vector $\vec x_d$ is known for each arm $d$, as it is computed from simple lexical query-document similarities as shown in Eq. \ref{eq:features}. 
We assume there is an unknown weight vector $\vec{w}$ such that the expected reward of document $d$ (i.e., relevance score from the reranker) is approximately $\vec{w}^{\top}\vec x_d$. This is the standard linear bandit assumption, and it implies that our linear surrogate model is capable of predicting relevance if we find the right weights.
The challenge is to learn $\vec{w}$ by sampling a limited number of arms (documents to send to the reranker) and observing their rewards (reranker scores). This is akin to a stochastic linear bandit problem \cite{reda2021top}, where each teacher evaluation provides a noisy signal of the true relevance.

To learn $\vec{w}$, we define a loss function that quantifies the error in our surrogate’s predictions. Whenever the teacher $\psi$ evaluates a document $d$ for query $Q$, we get a relevance score or label $y = \psi(Q,d)$ (e.g., a relevance score from MonoT5). We want our surrogate to accurately estimate this value. For a given weight vector $\vec{w}$, the estimation error for document $d$ can be written as:
\begin{equation}
\label{eq:loss}
\mathcal{E}(\vec{w}; Q, d, \vec x_d) = \frac{1}{2}\Big( \psi(Q,d) -s(d;\vec{w}) \Big)^2~
\end{equation}
The surrogate model’s goal is to find weights $\vec{w}$ that minimize this error, using linear least squares, for the documents that have been evaluated by the teacher so far.
This setup draws inspiration from the linear bandit \cite{purohit2025sample} and online regression that has also been used in document ranking recently~\cite{reda2021top,ore}. 
In particular, since each document (arm) can only be evaluated once by the teacher, our setting can be seen as an instance of \textbf{disposable bandits} \cite{disposable_bandits}, where each arm is sampled only once. 
We also impose a strict overall budget $c$ on the total number of teacher calls, to make the learning process sample-efficient and hence our approach can be seen as \textbf{budgeted disposable bandits}.

\subsection{Budget-aware Online Optimization}
Once we have the candidate pool $\mathcal{C}$ and the surrogate model structure in place, the next step is to select which documents to send to the teacher for evaluation, under the constraint that we can only afford $B$ teacher evaluations in each round, with $c$ being the total budget. Our strategy is an iterative, feedback-driven selection that uses the estimated relevance from the surrogate model to decide which documents (and implicitly reformulations) to focus on. 


Algorithm~\ref{algo:cerberus} implements the core online relevance estimation loop of \approach{}, which adaptively selects documents for expensive reranking under a strict budget. 
Starting from the original query $Q$, the algorithm first generates a set of reformulated queries $\mathcal{Q}$ (\textbf{Step 2}) and retrieves a candidate pool $\mathcal{C}$ (\textbf{Steps 3,4}) by issuing each reformulation independently to a first-stage retriever and combining the results with the initial retrieval $R_0$ pool from the original query (\textbf{Step 6}). 
Each document $d \in \mathcal{C}$ is represented by a feature vector $\vec{x}_d$ encoding its relevance signals with respect to the reformulations and the original query. A linear surrogate model with weights $\mathbf{w}_t$ estimates the utility of each document via $s(d;\mathbf{w}_t)=\mathbf{w}_t^\top\vec{x}_d$ (\textbf{Step 10}). At each iteration, the algorithm selects a small batch of documents with the highest estimated utility, subject to the remaining reranking budget $c$, and queries a strong teacher reranker $\psi$ (MonoT5) to obtain relevance scores anchored to the original query (\textbf{Steps 13-15}). These teacher scores are then used to update the surrogate weights by minimizing a squared prediction error, effectively performing online feature selection over reformulations (\textbf{Step 16}). Scored documents are removed from the candidate pool and added to the result set (\textbf{Steps 18,19}), and the process repeats until the budget is exhausted. Through this iterative loop, \approach{} concentrates the reranking effort on documents retrieved by the most effective reformulations, thereby increasing recall while actively suppressing query drift.

\algdef{SE}[DOWHILE]{Do}{doWhile}{\algorithmicdo}[1]{\algorithmicwhile\ #1}

\begin{figure}[t]
\begin{algorithm}[H]
\caption{ The \approach{} algorithm}
\begin{algorithmic}[1]
\Require Original query $Q$, initial retrieved pool $R_0$, batch size $b$, reranking budget $c$, 
feature vector $\vec{x}_d$ for each document $d$
\Ensure Scored result set $R_1$

\State $R_1 \gets \emptyset$ \Comment{Documents scored by the teacher}

\State $\mathcal{Q} \gets \textsc{Reformulate}(Q)$
\Comment{Generate reformulated queries}

\ForAll{$Q_i \in \mathcal{Q}$}
    \State $D_i \gets \text{BM25}(Q_i)$
\EndFor

\State $\mathcal{C} \gets R_0 \cup \bigcup_{i} D_i$
\Comment{Candidate pool (documents as arms)}

\State Initialize $\vec{w}_1 \sim \mathcal{N}(\mathbf{0}, \mathbf{1})$, $t \gets 1$
\Comment{Surrogate weights and iteration counter}

\Do
    \State \textbf{for all} $d \in \mathcal{C}$:
    \State \hspace{1em} $s(d;\vec{w}_t) \gets \vec{w}_t^\top \vec{x}_d$
    \Comment{Estimated utility}

    \State $B \gets$ top-$b$ documents from $\mathcal{C}$ by $s(d;\vec{w}_t)$
    \State $B \gets B$ truncated to respect remaining budget $c$
    
    \If{$|B| > 0$}
        \State \textbf{for all} $d \in B$:
        \State \hspace{1em} $y_d \gets \psi(Q,d)$
        \Comment{Teacher reranking (MonoT5)}

        \State Update $\vec{w}_{t+1}$ by minimizing
        \[
            \sum_{d \in B} \tfrac{1}{2}\big(\psi(Q,d) - \vec{w_t}^\top \vec{x}_d\big)^2
        \]
        \Comment{Using linear least-squares optimization}
    \EndIf

    \State $R_1 \gets R_1 \cup \{(d,y_d): \forall d \in B\}$
    \State $\mathcal{C} \gets \mathcal{C} \setminus B$
    \State $t \gets t+1$

\doWhile{$|R_1| < c$ and $\mathcal{C} \neq \emptyset$}

\end{algorithmic}
\label{algo:cerberus}
\end{algorithm}
\vspace{-1cm}
\end{figure}
By concentrating the limited teacher evaluations on the most promising documents (as predicted by the surrogate), this approach ensures that we find as many relevant documents as possible (high recall) while spending very few evaluations on irrelevant ones. Importantly, because the teacher always evaluates documents with respect to the original query $Q$, it provides a \textit{stable reference point} that prevents the system from drifting into evaluating documents that might only be relevant to a misconstrued reformulation.

During this process, features corresponding to reformulations that lead to many low-quality documents will be recognized and down-weighted early. For example, if $Q_j$ is a reformulation that retrieves mostly irrelevant documents, those documents (when eventually selected for reranker feedback) will receive low $y_d$ scores, and the weight $w_j$ corresponding to $Q_j$’s BM25 feature will decrease, because the surrogate will learn that a high BM25 score on $Q_j$ doesn’t correlate with true relevance. Conversely, a reformulation that consistently retrieves relevant documents will achieve high teacher scores for those documents, pushing its weight upward. This adaptive reweighting is done implicitly through the updates to surrogate.


\subsubsection{Efficiency considerations}
Another important aspect is efficiency: by focusing on top-scoring candidates first, we maximize the chance of finding relevant documents early. If the budget $c$ is much smaller than $|\mathcal{C}|$, the surrogate’s guidance ensures we pick the likely-best candidates rather than randomly sampling the pool. 
\section{Experimental Setup}
We aim to answer the following research questions:





\noindent \textbf{RQ1}: How effective is adaptive prioritization and ranking with \approach{} compared to existing reformulation methods?

\noindent \textbf{RQ2}: How robust is \approach{} to query drift as the number of reformulations increases?

\noindent \textbf{RQ3}: Under a fixed inference budget, is it more effective to allocate LLM capacity to upstream query reformulation with adaptive optimization (\approach{}) or to downstream document reranking?

\subsection{Datasets}


We evaluate on the MSMARCO passage corpus~\cite{bajaj2016ms} and validate on the TREC Deep Learning
(DL19--DL22) benchmarks~\cite{craswell2021trec}.
Specifically, MSMARCO passage-v1 is used with DL19 and DL20, while the
de-duplicated MSMARCO passage-v2 corpus is used with DL21 and DL22, along with
their corresponding de-duplicated qrels.
All indices are built using Terrier~\cite{ounis2005terrier}.
We report ranking effectiveness using nDCG@c and retrieval effectiveness using
Recall@c under re-ranking budgets $c \in \{50,100\}$.

\subsection{Ranking and Reformulation Models}

We use BM25 as the initial retriever via PyTerrier~\cite{macdonald2021pyterrier} and MonoT5-base as the cross-encoder ranker, fine-tuned on MSMARCO for all baselines and our approach.
For query reformulation, we use \texttt{gpt-4o-mini} in the main experiments (in Tables~\ref{tab:main} and~\ref{tab:msmarcov2}) and additionally evaluate open-source LLMs of varying scales, including Qwen2.5
0.5B and 32B~\cite{qwen2025qwen25technicalreport}. All experiments were carried out on 4 NVIDIA GeForce RTX 2080Ti GPUs.

\subsection{Baselines}
\subsubsection{Query Reformulation baselines} 

We compare \approach{} against representative training-free generative
reformulation methods spanning multiple categories:
(i) rephrasing-based methods like GenQR \cite{genqr}, and GenQREnsemble (GEns) \cite{genqr_ensemble},
(ii) pseudo-document generation -- Query2Doc \cite{wang2023query2doc},
(iii) question--answer-based reformulation -- QA-Expand \cite{seo2025qa},
and (iv) pseudo-relevance feedback methods --RM3 and GenQREnsemble+PRF (GEns + PRF).

For each baseline, we follow the combination strategies recommended in the
original work.
Unless otherwise stated, the number of reformulations is fixed to $n=5$, and
each reformulated query / combined query retrieves up to 100 candidates for fair comparison.
All reformulation pipelines are implemented using QueryGym~\cite{bigdeli2025querygymtoolkitreproduciblellmbased}.

\subsubsection{Exhaustive} We also use MonoT5 as a retriever on the MSMARCO passage corpus, scoring all documents exhaustively for a query denoted by ``Exhaustive Retrieval" in the tables in Section \ref{sec:experiments}.

\subsection{ Hyperparameters}
We use a temperature of 0.5 for reformulator LLMs to ensure diversity in generated reformulations while balancing quality, as lower temperatures lead to redundant queries. We do not use a higher temperature as this leads to more noisy reformulations due to hallucinations. While the default max\_tokens for all approaches is set to 256, for query2doc, we increase this to 800 as the approach is expected to generate a longer pseudo-document compared to other approaches. In \approach{} we set $|S|=15$, which is the number of top documents to consider from the already ranked pool for computing RM3 features. We use DL19 set as a validation set for tuning hyperparameters and DL20, DL21, and DL22 as test sets. For $RM3$, we set \textit{fb\_docs} to 5 and \textit{fb\_terms} to 10, and the \textit{original\_query\_weight} to 0.3. The weights of \approach{}  are initialized as described in Algorithm \ref{algo:cerberus}.

\section{Results and Analysis}
\label{sec:experiments}

\begin{table*}[t]
\centering
\caption{Effectiveness comparison of \approach{} across reformulation methods on
TREC DL19 and DL20. We report nDCG@c and Recall@c. We omit nDCG@10 our focus is more on mitigating drift to capture all relevant documents. Additionally, prior works find that nDCG@10 value saturates quickly during re-ranking~\cite{plaid_repro}. Significant improvements using paired t-test, $p<0.05$, with Bonferroni correction, over respective baselines marked with corresponding letters. Note that subscript: (F) generally denotes significance of \approach{} over the RRF variant. Best scores per block are in bold.}
\vspace{-0.8em}
{\small
\setlength{\tabcolsep}{6pt}
\begin{tabular}{lcccc|cccc}
\toprule
 & \multicolumn{4}{c|}{\textbf{DL19}} & \multicolumn{4}{c}{\textbf{DL20}} \\
Pipeline
& \multicolumn{2}{c|}{$c=50$}
& \multicolumn{2}{c|}{$c=100$}
& \multicolumn{2}{c|}{$c=50$}
& \multicolumn{2}{c}{$c=100$} \\
\cmidrule(lr){2-5}\cmidrule(lr){6-9}
 & nDCG@c & Recall@c & nDCG@c & Recall@c
 & nDCG@c & Recall@c & nDCG@c & Recall@c \\
\midrule

\multicolumn{9}{l}{\bf\textsc{Exhaustive Retrieval}} \\

MonoT5
& 0.625 & 0.512 & 0.611 & 0.599
& 0.592 & 0.576 & 0.593 & 0.670 \\

\midrule

BM25>>MonoT5 [B]
& 0.541 & 0.389 & 0.563 & 0.488
& 0.559 & 0.478 & 0.581 & 0.584 \\

\idnt w/ RM3
& 0.557 & 0.405 & 0.572 & 0.507
& 0.583 & 0.510 & 0.608 & 0.629 \\

\midrule
\multicolumn{9}{l}{\bf\textsc{Query Reformulation}} \\
\midrule

GenQR (G)
& 0.553 & 0.401 & 0.574 & 0.510
& 0.583 & 0.525 & 0.610 & 0.638 \\

GenQR + RRF [F]
& 0.547 & 0.394 & 0.569 & 0.505
& 0.582 & 0.522 & 0.617 & 0.644 \\

\rowcolor{reformoptgray}
GenQR + \approach{}
& ${}^{G}_{BF}$\textbf{0.600} & ${}^{G}_{BF}$\textbf{0.453}
& \bf ${}^{G}_{BF}$0.617 & ${}^{G}_{BF}$\textbf{0.561}
& \bf ${}^{G}_{BF}$0.634 & ${}^{G}_{BF}$\textbf{0.594}
& ${}^{G}_{BF}$\bf0.643 & ${}^{G}_{B}$\textbf{0.675} \\

\midrule

GEns (GE)
& 0.563 & 0.423 & 0.587 & 0.537
& 0.586 & 0.538 & 0.611 & 0.649 \\

GEns + RRF [F]
& 0.548 & 0.421 & 0.575 & 0.525
& 0.568 & 0.519 & 0.604 & 0.645 \\

\rowcolor{reformoptgray}
GEns + \approach{}
& \bf ${}^{GE}_{BF}$0.622 & ${}^{GE}_{BF}$\textbf{0.487}
& ${}^{GE}_{BF}$\textbf{0.634} & ${}^{}_{BF}$\textbf{0.584}
& ${}^{GE}_{BF}$\textbf{0.627} & ${}^{GE}_{BF}$\textbf{0.591}
& ${}^{GE}_{BF}$\bf0.641 & ${}^{}_{BF}$\textbf{0.671} \\

\midrule

GEns + PRF [R]
& 0.547 & 0.412 & 0.564 & 0.508
& 0.569 & 0.520 & 0.601 & 0.625 \\

\rowcolor{reformoptgray}
GEns + PRF + \approach{}
& ${}^{R}_{B}$\textbf{0.592} & ${}^{R}_{B}$\textbf{0.454}
& ${}^{R}_{B}$\textbf{0.626} & ${}^{R}_{B}$\textbf{0.575}
& ${}^{R}_{B}$\textbf{0.628} & ${}^{R}_{B}$\textbf{0.572}
& ${}^{R}_{B}$\textbf{0.637} & ${}^{R}_{B}$\textbf{0.671} \\

\midrule

QA-Expand (Q)
& 0.613 & 0.468 & 0.633 & 0.575
& 0.606 & 0.576 & 0.622 & 0.669 \\

QA-Expand + RRF [F]
& 0.616 & 0.464 & 0.626 & 0.582
& 0.608 & 0.558 & 0.634 & 0.689 \\

\rowcolor{reformoptgray}
QA-Expand + \approach{}
& ${}^{Q}_{BF}$\textbf{0.672} & ${}^{Q}_{BF}$\textbf{0.526}
& ${}^{Q}_{BF}$\bf0.668 & ${}^{Q}_{BF}$\textbf{0.629}
& ${}^{Q}_{BF}$\textbf{0.651} & ${}^{}_{BF}$ \textbf{0.615}
& ${}^{Q}_{BF}$\bf0.661 & ${}^{Q}_{BF}$\textbf{0.714} \\

\midrule

Query2Doc (QD)
& 0.627 & 0.502 & 0.630 & 0.605
& 0.632 & 0.596 & 0.649 & 0.688 \\

\rowcolor{reformoptgray}
Query2Doc + \approach{}
& ${}^{QD}_{B}$\textbf{0.671} & ${}^{}_{B}$ \textbf{0.538}
& ${}^{QD}_{B}$\textbf{0.673} & ${}^{}_{B}$ \textbf{0.639}
& ${}^{}_{B}$ \textbf{0.662} & ${}^{}_{B}$\textbf{0.616}
& ${}^{QD}_{B}$\textbf{0.680} & ${}^{}_{B}$ \textbf{0.730} \\

\bottomrule
\end{tabular}
}
\vspace{-0.5em}
    \label{tab:main}
\end{table*}


\begin{table*}[t]
\centering
\caption{Effectiveness comparison of \approach{} across reformulation methods on TREC DL21 and DL22. We report nDCG@c and Recall@c only. Significant improvements using paired t-test, $p<0.05$, with Bonferroni correction, over respective baselines marked with corresponding letters. Best scores per block are in bold.}
\vspace{-0.8em}
{\small
\setlength{\tabcolsep}{6pt}
\begin{tabular}{lcccc|cccc}
\toprule
 & \multicolumn{4}{c|}{\textbf{DL21}} & \multicolumn{4}{c}{\textbf{DL22}} \\
Pipeline
& \multicolumn{2}{c|}{$c=50$}
& \multicolumn{2}{c|}{$c=100$}
& \multicolumn{2}{c|}{$c=50$}
& \multicolumn{2}{c}{$c=100$} \\
\cmidrule(lr){2-5}\cmidrule(lr){6-9}
 & nDCG@c & Recall@c & nDCG@c & Recall@c
 & nDCG@c & Recall@c & nDCG@c & Recall@c \\
\midrule

\multicolumn{9}{l}{\bf\textsc{Baseline Retrieval}} \\

BM25>>MonoT5 [B]
& 0.436 & 0.242 & 0.433 & 0.331
& 0.290 & 0.115 & 0.275 & 0.164 \\

\idnt w/ RM3
& 0.455 & 0.274 & 0.457 & 0.375
& 0.287 & 0.115 & 0.275 & 0.161 \\

\midrule
\multicolumn{9}{l}{\bf\textsc{Query Reformulation}} \\
\midrule

GenQR (G)
& 0.382 & 0.224 & 0.398 & 0.318
& 0.292 & 0.136 & 0.288 & 0.193 \\

GenQR + RRF [F]
& 0.390 & 0.226 & 0.407 & 0.324
& 0.291 & 0.137 & 0.294 & 0.194 \\

\rowcolor{reformoptgray}
GenQR + \approach{}
& ${}^{G}_{BF}$\textbf{0.482} & ${}^{G}_{BF}$\textbf{0.299}
& ${}^{G}_{BF}$\textbf{0.488} & ${}^{G}_{BF}$\textbf{0.403}
& ${}^{G}_{BF}$\textbf{0.352} & ${}^{G}_{BF}$\textbf{0.170}
& ${}^{G}_{BF}$\textbf{0.340} & ${}^{G}_{BF}$\textbf{0.236} \\

\midrule

GEns (GE)
& 0.421 & 0.266 & 0.432 & 0.370
& 0.289 & 0.135 & 0.288 & 0.189 \\

GEns + RRF [F]
& 0.390 & 0.232 & 0.424 & 0.350
& 0.274 & 0.125 & 0.272 & 0.176 \\

\rowcolor{reformoptgray}
GEns + \approach{}
& ${}^{GE}_{BF}$\textbf{0.480} & ${}^{GE}_{BF}$\textbf{0.308}
& ${}^{GE}_{BF}$\textbf{0.484} & ${}^{GE}_{BF}$\textbf{0.402}
& ${}^{GE}_{BF}$\textbf{0.351} & ${}^{GE}_{BF}$\textbf{0.171}
& ${}^{GE}_{BF}$\textbf{0.334} & ${}^{GE}_{BF}$\textbf{0.227} \\

\midrule

GEns + PRF [R]
& 0.408 & 0.251 & 0.410 & 0.332
& 0.251 & 0.100 & 0.248 & 0.145 \\

\rowcolor{reformoptgray}
GEns + PRF + \approach{}
& ${}^{R}_{B}$\textbf{0.491} & ${}^{R}_{B}$\textbf{0.315}
& ${}^{R}_{B}$\textbf{0.485} & ${}^{R}_{B}$\textbf{0.409}
& ${}^{R}_{B}$\textbf{0.317} & ${}^{R}_{B}$\textbf{0.142}
& ${}^{R}_{B}$\textbf{0.322} & ${}^{R}_{B}$\textbf{0.213} \\

\midrule

QA-Expand (Q)
& 0.483 & 0.329 & 0.482 & 0.437
& 0.349 & 0.186 & 0.345 & 0.265 \\

QA-Expand + RRF [F]
& 0.498 & 0.328 & 0.493 & 0.441
& 0.361 & 0.191 & 0.342 & 0.254 \\

\rowcolor{reformoptgray}
QA-Expand + \approach{}
& ${}^{Q}_{BF}$\textbf{0.538} & ${}^{}_{BF}$\textbf{0.356}
& ${}^{Q}_{BF}$\textbf{0.523} & ${}^{}_{B}$\textbf{0.463}
& ${}^{Q}_{BF}$\textbf{0.403} & ${}^{Q}_{BF}$\textbf{0.214}
& ${}^{}_{BF}$\textbf{0.378} & ${}^{}_{BF}$\textbf{0.288} \\

\midrule

Query2Doc (QD)
& 0.456 & 0.316 & 0.466 & 0.420
& 0.387 & 0.229 & 0.366 & 0.306 \\

\rowcolor{reformoptgray}
Query2Doc + \approach{}
& ${}^{QD}_{B}$\textbf{0.531} & ${}^{}_{B}$ \textbf{0.354}
& ${}^{QD}_{B}$\textbf{0.533} & ${}^{QD}_{B}$ \textbf{0.481}
& ${}^{QD}_{B}$\textbf{0.425} & ${}^{}_{B}$ \textbf{0.240}
& ${}^{QD}_{B}$\textbf{0.410} & ${}^{}_{B}$ \textbf{0.325} \\

\bottomrule
\end{tabular}
}
\vspace{-1.em}
    \label{tab:msmarcov2}
\end{table*}

\subsection{Query Reformulation based Ranking Effectiveness of \approach{}}

To answer \textbf{RQ1}, we compare how query reformulation approaches augmented with \approach{} affect ranking performance. The results are shown on TREC-DL19 and DL20 in Table \ref{tab:main} and on TREC-DL21 and DL22 in Table \ref{tab:msmarcov2}. We observe that all query reformulation approaches augmented with \approach{} outperform the classical retrieve-rerank pipeline BM25>>MonoT5. For instance, on DL22 GenQR + \approach{} outperforms BM25>>MonoT5 by \textbf{43.90\%} at budget $c=100$ in Recall@c. We also observe that query reformulation approaches augmented with \approach{} outperform the pseudo relevance feedback based BM25+RM3 >>MonoT5.
We also observe from Table \ref{tab:main}, that some of the query reformulation approaches, when augmented with \approach{}, outperform the exhaustive retrieval approach, which scores all documents in the corpus using the expensive re-ranker without any first-stage retrieval.  




Finally, comparing the effectiveness of \approach{} with respect to each query reformulation approach. We observe that GenQR, which generates multiple variants of the original query by rephrasing, has significant drift with respect to the semantics of the original query, offering only marginal gains over the retrieve-rerank pipeline (with base query). We observe that in general this primarily occurs because some of the reformulated queries drift from the semantics of the original query, causing deterioration of the candidate document pool retrieved for ranking. This demonstrates the need for prioritizing query reformulations adaptively reduce drift to enhance downstream ranking performance. 

Augmenting GenQR with \approach{}, significantly reduces drift, as observed by Recall@50 on DL21, where performance improves by \textbf{33.48\%} when compared to GenQR and by \textbf{32.3\%} compared to the RRF variant of GenQR. \approach{} jointly determines the reformulations to prioritize through learned weights and to prioritize the most promising document for ranking through estimated utilities. 

In addition to prioritizing queries, we posit that our approach also enables prioritizing key documents from the pool, balancing exploration and exploitation. Our joint optimization setup leverages BM25 scores of reformulated queries and documents, combined with RM3 features obtained using documents already deemed relevant by the reranker. Here, \approach{} could potentially \textit{exploit} documents with high relevance (BM25) scores with respect to reformulated queries or could decide to \textit{explore} documents with high lexical similarity to terms from documents already deemed relevant.

Note that the candidate documents pool in RRF variants is similar to the \approach{}, but RRF considers all reformulations to be equally important when performing rank fusion. Although this is intended to capture multiple facets of the query, it often aggregates across redundant or low-quality candidates as also observed in prior work \cite{seo2025qa} resulting in degradation of performance. 

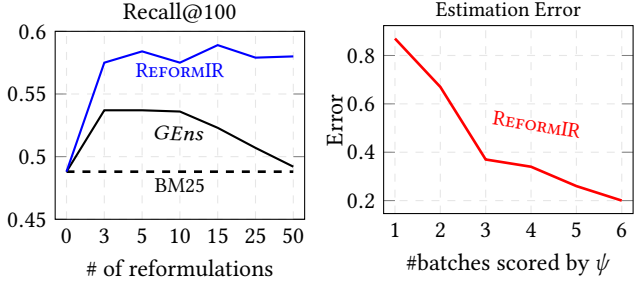
\begin{figure}[]
     \begin{subfigure}{0.48\columnwidth}
         \centering
 \begin{tikzpicture}
		\begin{axis}[
			width =  1.2\linewidth,
			height =1\linewidth,
			major x tick style = transparent,
			grid = major,
		    grid style = {dashed, gray!20},
			xlabel = {\# of reformulations},
			ylabel = {},
			title={Recall@100},
            title style={yshift=-1.5ex}, 
            symbolic x coords={0,3,5,10,15,25,50},
            xtick={0,3,5,10,15,25,50},
            xtick distance=20,
            enlarge x limits=0.05,
            xlabel near ticks,
            ylabel near ticks,
            ymin=0.45,
            ymax=0.6,
            every axis y label/.style={at={(-0.18, 0.5)},rotate=90,anchor=near ticklabel},
			]

	\addplot [color=black, style= dashed, line width = 1.0pt] table [x index=0, y index=1, col sep = comma] {fig/genqr_ensemble_varying_n.txt}
         node[pos=0.5, below, yshift=1pt] {\small{BM25}};
			

   	\addplot [color=black, line width = 0.8pt] table [x index=0, y index=3, col sep = comma] {fig/genqr_ensemble_varying_n.txt}
    node[pos=0.5,yshift=-2pt, sloped, below] {\small{$GEns$}};


   	\addplot [color=blue, line width = 0.8pt] table [x index=0, y index=4, col sep = comma] {fig/genqr_ensemble_varying_n.txt}
    node[pos=0.5, yshift=3pt, below] { \small{\approach{}}};
   
    \end{axis}
    \end{tikzpicture}    
    \label{fig:genqr_ensemble}
 \end{subfigure}
     \begin{subfigure}{0.48\columnwidth}

     \end{subfigure}
   \begin{subfigure}{0.48\columnwidth}
        \centering
    \begin{tikzpicture}
		\begin{axis}[
			width =  1.2\linewidth,
			height =1.0\linewidth,
			major x tick style = transparent,
			grid = major,
		    grid style = {dashed, gray!20},
			xlabel = {\#batches scored by $\psi$},
			ylabel = {Error},
			title={\small{Estimation Error}},
            title style={yshift=-1.5ex}, 
            symbolic x coords={1,2,3,4,5,6,7},
            xtick={1,2,3,4,5,6,7},
            xtick distance=20,
            enlarge x limits=0.05,
            enlarge y limits=0.08,
            xlabel near ticks,
            ylabel near ticks,
            every axis y label/.style={at={(-0.13, 0.5)},rotate=90,anchor=near ticklabel},
			]

   	\addplot [color=red, line width = 1.0pt] table [x index=0, y index=4, col sep = comma] {fig/error_plot.txt}
    node[pos=0.6, yshift=10pt, sloped, above] {\small{\approach{}}};
   
        \end{axis}
    \end{tikzpicture}    
    \label{fig:hybrid_error}
    \end{subfigure} 

\vspace{-1em}
\caption{Recall when the number of reformulations increases $GEns$ and estimation error (right) when the number of batches of documents scored by cross-encoder ($\psi$) varies for \approach{} comparison on the \textbf{TREC DL19} dataset for ranking budget of 100 and batch of size 16.}
      \label{fig:scaling_exp}  
    \Description{}

\vspace{-1em}
\end{figure}

While QA-Expand generates focused question-answer pairs to mitigate drift, the generated questions could potentially focus on implicit facets of the query that are not critical, resulting in some of the questions and answers being irrelevant to the original query. However, as we observe from results in Tables \ref{tab:main} and \ref{tab:msmarcov2}, \approach{} helps adaptively prioritize the most relevant queries (question-answer pairs) and documents, leading to gains of \textbf{15.05\%} for Recall@50 on DL22 and upto \textbf{12.39\%} on DL19.

We also provide a concrete example of drift mitigation in QA-Expand by \approach{} from our manual analysis as shown in \textbf{Figure \ref{fig:qual_anaysis}} (Note: here reformulations are QA pairs but we omit answers for readability and also due to space constraints). In this example, we observe that reformulated query 1 has a high individual recall compared to queries 2 and 3. This is primarily because reformulated query 3, which poses a question about the purpose of the ingredients, drifts from the core semantics of the query, which concerns regarding the list of ingredients of \texttt{supartz injections} and not their purpose. We also observe that reformulated query 1 is a better phrasing of the ambiguous query compared to query 2. We observe that the learned weights in \approach{} weigh query 1 are highly relative to the other two reformulations, demonstrating the drift mitigation. This results in a final recall of \textbf{0.36} compared to baseline, which has a lower recall of \textbf{0.26} for this query.

\usepgfplotslibrary{colorbrewer} 
\usetikzlibrary{patterns}

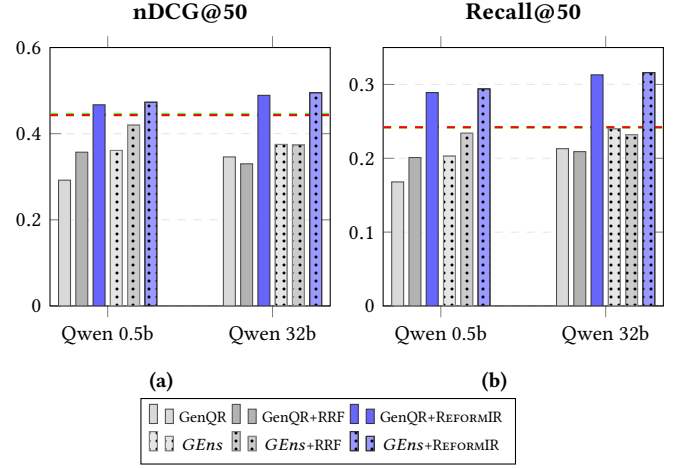
\begin{figure}[t]
        \begin{subfigure}{0.48\columnwidth}
    \begin{tikzpicture}
        \pgfplotsset{
            base plot style/.style={
                ybar,
                width=1.3\columnwidth, 
                height=5.cm,
                bar width=4.5pt, 
                symbolic x coords={ Qwen-0.5b, Qwen-32b},
                xtick={ Qwen-0.5b, Qwen-32b},
                enlarge x limits=0.35,
                ymajorgrids=true,
                grid style={dashed, gray!20},
                ylabel style={font=\small},
                tick label style={font=\small},
                legend style={font=\scriptsize}, 
                cycle list={
                    {fill=gray!30, draw=black!70},   
                    {fill=gray!60, draw=black!70},   
                    {fill=blue!60, draw=black!80},   
                    {fill=gray!20, draw=black!50, postaction={pattern=dots}}, 
                    {fill=gray!40, draw=black!50, postaction={pattern=dots}}, 
                    {fill=blue!40, draw=black!80, postaction={pattern=dots}}  
                }
            }
        }

        \begin{axis}[
            name=plot1,
            base plot style,
            ylabel={},
            ymin=0, ymax=0.6,
            title={\textbf{nDCG@50}},
            legend to name=sharedlegend,
            legend columns=3,
            xticklabels={ Qwen 0.5b, Qwen 32b},
        ]

\draw[green, dashed, thick] ({rel axis cs:0,0} |- {axis cs:Qwen-0.5b,0.445}) -- ({rel axis cs:1,0} |- {axis cs:Qwen-0.5b,0.445});

\draw[red, dashed, thick] ({rel axis cs:0,0} |- {axis cs:Qwen-0.5b,0.443}) -- ({rel axis cs:1,0} |- {axis cs:Qwen-0.5b,0.443});

                        \addplot coordinates { (Qwen-0.5b,0.292) (Qwen-32b,0.346)};
            \addplot coordinates { (Qwen-0.5b,0.357) (Qwen-32b,0.330)};
            \addplot coordinates { (Qwen-0.5b,0.467) (Qwen-32b,0.489)};
            \addplot coordinates { (Qwen-0.5b,0.361) (Qwen-32b,0.375)};
            \addplot coordinates { (Qwen-0.5b,0.420) (Qwen-32b,0.374)};
            \addplot coordinates { (Qwen-0.5b,0.473) (Qwen-32b,0.495) };

            \legend{GenQR, GenQR+RRF, GenQR+\approach{}, $GEns$, $GEns$+RRF, $GEns$+\approach{}}
            \end{axis}
\end{tikzpicture}
    \caption{}
\end{subfigure}
\hfill
    \begin{subfigure}{0.48\columnwidth}
\begin{tikzpicture}
            \pgfplotsset{
            base plot style/.style={
                ybar,
                width=1.3\columnwidth, 
                height=5.cm,
                bar width=4.5pt, 
                symbolic x coords={ Qwen-0.5b, Qwen-32b},
                xtick={ Qwen-0.5b, Qwen-32b},
                enlarge x limits=0.35,
                ymajorgrids=true,
                grid style={dashed, gray!20},
                ylabel style={font=\small},
                tick label style={font=\small},
                legend style={font=\scriptsize}, 
                cycle list={
                    {fill=gray!30, draw=black!70},   
                    {fill=gray!60, draw=black!70},   
                    {fill=blue!60, draw=black!80},   
                    {fill=gray!20, draw=black!50, postaction={pattern=dots}}, 
                    {fill=gray!40, draw=black!50, postaction={pattern=dots}}, 
                    {fill=blue!40, draw=black!80, postaction={pattern=dots}}  
                }
            }
        }
        \begin{axis}[
            name=plot2,
            at={(plot1.below south east)},
            anchor=above north east,
            yshift=-0.4cm, 
            base plot style,
            ylabel={},
            ymin=0, ymax=0.35,
            title={\textbf{Recall@50}},
            xticklabels={ Qwen 0.5b, Qwen 32b},
        ]

\draw[green, dashed, thick] ({rel axis cs:0,0} |- {axis cs:Qwen-0.5b,0.242}) -- ({rel axis cs:1,0} |- {axis cs:Qwen-0.5b,0.242});

\draw[red, dashed, thick] ({rel axis cs:0,0} |- {axis cs:Qwen-0.5b,0.242}) -- ({rel axis cs:1,0} |- {axis cs:Qwen-0.5b,0.242});

              \addplot coordinates { (Qwen-0.5b,0.168) (Qwen-32b,0.213)};
            \addplot coordinates { (Qwen-0.5b,0.201) (Qwen-32b,0.209)};
            \addplot coordinates { (Qwen-0.5b,0.289) (Qwen-32b,0.313)};
            \addplot coordinates { (Qwen-0.5b,0.203) (Qwen-32b,0.240)};
            \addplot coordinates { (Qwen-0.5b,0.234) (Qwen-32b,0.232)};
            \addplot coordinates { (Qwen-0.5b,0.294) (Qwen-32b,0.316)};
            

        \end{axis}
    \end{tikzpicture}
    \caption{}
    \end{subfigure}
    \vspace{5pt}
    \ref{sharedlegend}
    \caption{Impact of using LLMs of different sizes as query reformulators. The red dashed line (BM25>>RankLlama \& BM25>>RankZephyr) serves as a baseline and note that nDCG@50 and Recall@50 for both baselines have marginal difference causing overlap in plot. Blue bars represent \approach{}, while dotted patterns indicate ensemble variants.}
    \label{fig:results_comparison}
\end{figure}

Query2doc departs from other reformulation approaches by generating pseudo-documents that might potentially contain relevant terms, and the authors claim it does not rely \cite{wang2023query2doc} on initial retrieval results, which might be noisy or irrelevant. While the pseudo-documents contain additional context and contribute additional relevant terms, they may also sometimes contribute significantly to drift due to hallucination. We particularly observe this issue in DL21, where  Query2Doc + \approach{} offers statistically significant gains of upto \textbf{14.52\%} in Recall@100. It also achieves \textbf{16.45\%}, \textbf{14.38\%} gains in nDCG@50 and nDCG@100 respectively. This demonstrates that \approach{} can help automatically downweigh low-quality queries (pseudo-docs) that retrieve irrelevant documents when drift occurs with respect to the original query, with help of reranker feedback.

\textbf{Insight 1:} \textit{\approach{} serves as a training-free adapter mitigating drift in query reformulations, leading to gains of upto \textbf{33.48\%}. It achieves this through joint optimization over reformulations and documents from the candidate pool using iterative feedback} 

\subsection{Effect of \approach{} with scaling the number of  query reformulations}

\begin{figure}[t]
    \centering
    \begin{tikzpicture}
        \pgfplotsset{
            latency plot style/.style={
                ybar,
                width=.7\columnwidth,
                height=5.0cm,
                bar width=7pt,
                symbolic x coords={50,100},
                xtick=data,
                enlarge x limits=0.25,
                ymajorgrids=true,
                grid style={dashed, gray!20},
                ylabel style={font=\small},
                xlabel style={font=\small},
                tick label style={font=\small},
                legend style={
                    font=\footnotesize,
                    at={(0.5,-0.35)},
                    anchor=north,
                    legend columns=3
                },
                cycle list={
                    {fill=orange!70, draw=black!80}, 
                    {fill=gray!60,   draw=black!70}, 
                    {fill=blue!60,   draw=black!80}  
                }
            }
        }

        \begin{axis}[
            latency plot style,
            ymode=log,
            log basis y=10,
            ylabel={Mean latency per query (ms, log scale)},
            xlabel={Re-ranking budget ($c$)},
            title={\textbf{Re-ranking Latency vs Budget}},
            ymin=1e3, ymax=3e4,
            log ticks with fixed point
        ]

            \addplot+[
                error bars/.cd,
                y dir=both,
                y explicit
            ] coordinates {
                (50,10863.2)+- (0,0.56)
                (100,17562.6)+- (0,0.56)
            };

            \addplot+[
                error bars/.cd,
                y dir=both,
                y explicit
            ] coordinates {
                (50,3048.81)+- (0,0.71)
                (100,3583.7)+- (0,0.79)
            };

            \addplot+[
                error bars/.cd,
                y dir=both,
                y explicit
            ] coordinates {
                (50,3257.41)+- (0,0.76)
            (100,3867.43  ) +- (0,0.87)
            };

            \legend{RankLLaMA, $GEns$+RRF, $GEns$+\approach{}}

        \end{axis}
    \end{tikzpicture}

    \caption{Mean latency per query (log scale) at different re-ranking budgets. Qwen2.5 (0.5B) is used as a reformulator for $GEns$+RRF and $GEns$+\approach{} with MonoT5 as re-ranker. Std-dev is small, hence not visible. }
    \label{fig:latency_reranking}
    \vspace{-1em}
\end{figure}
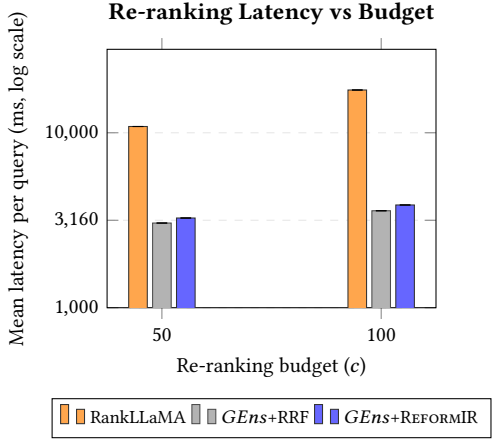

To answer \textbf{RQ2}, we vary the number of reformulations and observe the effect of drift on retrieval and ranking performance.  The results are as shown in Figures \ref{fig:genqr}, \ref{fig:scaling_exp} (left). We observe that in Figure \ref{fig:genqr}, GenQR gradually reduces in ranking performance with an increase in the number of reformulations $n=3,5,10,...,25, 50$. We observe that this could primarily result from drift in the reformulated queries, resulting in low-quality results from first-stage retrieval. Further, at a large number of reformulations $n=50$, the performance drops sharply, rendering the performance worse than the classical retrieve-rerank baseline (BM25>>MonoT5) using the base query, indicating that a significant number of reformulated queries cause drift. 

However, we observe that GenQR augmented with \approach{} leads to high and stable Recall@100 as observed in Figure \ref{fig:genqr}. We observe that this is primarily because our approach adaptively prioritizes the documents to consider through our simple surrogate model which is iteratively updated based on ranker feedback. From Figure \ref{fig:scaling_exp} (right), we observe that the surrogate's estimation error decreases as more feedback is sampled to update the surrogate. When computing the utility of a document, a weighted combination of BM25 scores with respect to each reformulation and related RM3 features are computed.  Hence, by consequence, these weights of the surrogate model updated using ranker feedback, also up-weighs or down-weighs the query reformulations based on their contribution to the document's utility and ranker feedback. Hence, query reformulations which drift from semantics of the original query as indicated by ranker feedback, and the corresponding documents which are irrelevant are not prioritized in \approach{}. This results in adaptive filtering of such low-quality queries and corresponding documents from the candidate pool, resulting in high recall and ranking performance. We observe similar results for other reformulation approaches like GEns in the presence of drift as observed in Figure \ref{fig:scaling_exp} (left). Hence, with an increasing number of reformulations, \approach{} provides an adaptive sample-efficient mechanism drift effectively compared to existing heuristic and rank fusion-based approaches.

\textbf{Insight 2}: \textit{Scaling the number of reformulations significantly degrades performance for classical reformulation approaches. However, when augmented with \approach{}, ranking performance remains stable and better than the classical variants.}

\subsection{Efficiency of \approach{} vs LLM based re-ranking and variation in \approach{} with generators of different scales}

To answer \textbf{RQ3}, we also vary the LLM used for reformulation and observe the impact on ranking performance across several reformulation approaches with and without \approach{}. We also compare \approach{} with MonoT5 re-ranker to  LLM based re-rankers like RankLlaMA without reformulation to analyze whether it is efficient and effective to use LLMs for reformulation vs LLMs for re-ranking.

\subsubsection{Effect of \approach{} with  generative models of different scales}

We first analyze the effect of \approach{} across two different reformulation approaches when open-source LLMs of different scales are employed.
The results can be observed in Figure \ref{fig:results_comparison}. We experiment with different open-source LLM - Qwen series with models of different scales - Qwen (0.5B and 32B) and with MonoT5 as the reranker. We observe that GEns suffers from a drop in performance at budgets $c=50$ and $c=100$. We observe that this is primarily because some reformulations from Qwen models comprise hallucinations when compared to \texttt{gpt-4o-mini}. Additionally, GEns ensembles reformulations from multiple variations of instructions to LLM, which might result in compounding semantic drift. However, we observe that GEns + \approach{} aids in mitigating drift, offering upto \textbf{44.83\%} gains in Recall@50 and \textbf{22.87\%} in Recall@100. 

We also perform experiments with one of the larger-scale Qwen models (32B) and observe little to no gains in the baseline performance of GenQR. However, we observe that when augmented with \approach{}, it aids in mitigating drift and leads to improvement in Recall@50 \textbf{46.95\%} and similar gains in nDCG@c ($c=50,100$) and Recall@100. However, the gains in Recall@c and nDCG@c of reformulation approaches (GenQR, GEns) augmented with \approach{} with reformulations from Qwen 32B are only marginally better than corresponding performance on reformulations from Qwen 0.5B. This demonstrates that primarily, performance gains of \approach{} stem from mitigating drift in query reformulations. It achieves this by adaptively prioritizing high-quality queries, and the relative gains are transferable to other open-source LLMs used for reformulation. Secondly, our results in Figure \ref{fig:results_comparison} demonstrate that augmenting reformulation approaches with \approach{} enables leveraging less expensive LLMs (0.5B scale) for reformulation.

\subsubsection{Efficiency of \approach{}}
We also analyze the efficiency and overhead incurred by \approach{} when compared to classical reformulation approaches. The results are as shown in Figure \ref{fig:latency_reranking}. We observe that \approach{} only adds marginal overhead over $GEns+RRF$, demonstrating that the adaptive optimization contributes to only minor overheads in terms of latency. Additionally, we also compare \approach{} to employing the RankLlaMA (7B) as a re-ranker without any reformulation to discern if it is more efficient and effective to use expensive LLMs in the ranking stage or employ LLMs (like Qwen2.5-0.5B) for the reformulation stage coupled with \approach{}. From Figure \ref{fig:results_comparison}, we observe that RankLlaMA has worse ranking performance when compared to different reformulation approaches augmented with \approach{}, which uses only MonoT5 - a relatively inexpensive ranker compared to RankLlaMA.
We also observe from Figure \ref{fig:latency_reranking} that \approach{} is \textbf{3.3x-4.5x} more efficient than RankLlaMA based re-ranking as it employs an inexpensive MonoT5 as a ranker and can provide high-quality reformulations even with a LLM of modest scale, as shown in Figure \ref{fig:results_comparison}. 

\textbf{Insights}: \textit{(1) Across diverse reformulation generators of different scales, \approach{} consistently improves performance. 
2)Augmenting reformulation approaches with \approach{} only adds marginal latency cost compared to classical approaches and is still \textbf{3.3x-4.5x} more efficient than LLM based re-ranking.
3) Hence, it is more efficient to employ LLMs upstream for reformulation coupled with adaptive optimization of \approach{} than for downstream reranking.
 }
\section{Conclusion}

In this paper, we introduce \approach{} as a simple adaptive optimization approach which can reduce drift in different query reformulation approaches. We achieve this by using the relevance of documents to original query as reference point and estimate this using an inexpensive linear surrogate which employs query reformulations as features by using lexical similarities between reformulation and documents. The estimated utilities are used to adaptively sample reranker feedback and update the surrogate. This enables joint optimization for prioritization of query reformulations and documents. We demonstrate that \approach{} achieves gains in recall mitigating drift. In future, we plan to explore further on updating the reformulators online using ranker feedback and exploring even smaller LLMs (<0.5B) to further reduce latency.

\section*{Acknowledgments}
This work was partially funded by the Bundesministerium für Wirtschaft und Energie (BMWE), Germany, in the context of the 8ra Initiative ("Soofi", 13IPC040E), and by the European Union’s Horizon Europe Research and Innovation Programme JustREACH under Grant Agreement No 101214666. The experiments were enabled by resources provided by the National Academic Infrastructure for Supercomputing in Sweden (NAISS), partially funded by the Swedish Research Council through grant agreement no. 2022-06725 and Zeus cluster provided by Department of Computer and Systems Sciences at Stockholm University.

\bibliographystyle{ACM-Reference-Format}
\balance
\bibliography{bib}

\end{document}